\documentclass[10pt,english]{revtex4}
\usepackage[T1]{fontenc}
\usepackage[latin9]{inputenc}
\usepackage{amsmath}
\usepackage{graphicx}
\usepackage{esint}
\usepackage{verbatim}
\makeatletter
\makeatletter
\usepackage{babel}

\makeatother

\usepackage{babel}
\begin{document}

\title{The tunneling  decay rate in QFT  beyond the thin wall approximation}

\author{G. Flores-Hidalgo}
\email{gfloreshidalgo@unifei.edu.br}
\affiliation{Instituto de F\'{\i}sica e Qu\'{\i}mica, Universidade Federal de Itajub\'a, 37500-903, MG,  Brazil.}

\begin{abstract}
The tunneling decay rate per unit volume in Quantum Field Theory (QFT), at order  $\hbar$, is given by $\Gamma/V = Ae^{-B}$, where $B$ is the Euclidean action evaluated at the so-called bounce, and $A$ is proportional to the determinant of a second-order differential operator. The dominant contribution comes from the exponential factor. To estimate $\Gamma/V$, one must determine the bounce configuration, which satisfies a highly nonlinear equation. A common approach in the literature is the thin-wall approximation. In this work, we extend the formalism to cases where the thin-wall approximation is not valid. We employ a simple variational method to estimate both the bounce and the decay rate, and we find good agreement between our results and full numerical calculations.
\end{abstract}
\maketitle

{\it Introduction:}
 The phenomenon of quantum tunneling in field theory is a long-standing and fundamental problem in theoretical physics, with broad 
implications across multiple domains. Originally introduced in the context of nuclear decay and quantum mechanics, tunneling in Quantum
 Field Theory (QFT) describes the decay of metastable vacuum states via nucleation of bubbles,  localized configurations that interpolate 
between a false and a true vacuum \cite{Coleman,Callan}. This mechanism underpins a variety of physical processes, from condensed matter systems 
\cite{Langer1967,Darbha}, to early-universe cosmology where scalar fields undergo first-order phase 
transitions \cite{Linde,Rubakov,Mark,Masina}, and high-energy physics where vacuum metastability plays a role in the fate of the Higgs sector \cite{Buttazzo2013,Andreassen2018}.
As briefly reviewed below, the mestatable vaccum decay rate per unit volume is exponentially suppressed by the Euclidean action of the bubble
 configurations. Understanding the precise form and dynamics of such tunneling events is essential to quantify rates of phase transitions, 
estimate lifetimes of metastable states, and explore their cosmological and phenomenological consequences.

Let us consider a real scalar field model, in
4-dimensional space-time, with Euclidean action given by
\begin{equation}
S_E(\varphi)=\int d^4 x_E \left[ \frac{1}{2}\left(\partial_{\mu}\varphi\right)^2 +
U(\varphi)\right]\;,
\label{action}
\end{equation}
where $U(\varphi)$ possesses an absolute minimum, denoted by $\varphi_-$, and at least one local (relative) minimum, denoted by $\varphi_+$. Classically, both $\varphi_+$ and $\varphi_-$ are stable configurations. However, at the quantum level, $\varphi_+$ becomes metastable, and the false vacuum associated with it can decay into the true vacuum corresponding to $\varphi_-$. This decay process can occur either at zero or finite temperature. 

At zero temperature, the study of false vacuum decay was initiated by Voloshin, Kobzarev, and Okun~\cite{Voloshin}, and later developed by Callan and Coleman~\cite{Coleman,Callan}, who introduced the so-called bounce method for analyzing quantum vacuum decay. In this context the decay rate per unit
volume $V$ is given by
\begin{equation} 
\frac{\Gamma}{V}=\left(
\frac{\Delta S_{E}}{2\pi}\right)^{2}
\left[\frac{{\rm det}'\left(-\partial_E^2+U''(\varphi_b)\right
)}{{\rm det}\left(-\partial^2_{E}+U''(\varphi_+)\right)}\right]^{-1/2}
e^{-B}\left(1+{\cal O}(\hbar)\right) \;,
\label{decayrate} 
\end{equation} 
where  $\partial^2_{E}=
\frac{\partial^2}{\partial{\tau^2}}+\frac{\partial^2}{\partial{x^2}}
+\frac{\partial^2}{\partial{y^2}}+\frac{\partial^2}{\partial{z^2}}$,
$U''(\varphi)=\frac{d^2U(\varphi)}{d\varphi^2}$ and  
$B =S_E (\varphi_b) -S_E(\varphi_+)$ is the difference in Euclidean action evaluated at the bounce $\varphi_b$ and the false vacuum $\varphi_+$.  The prime on the determinant indicates that zero modes are omitted.

 The bounce is  solution of the  
Euclidean field equation of motion, $\delta S_E/\delta 
\varphi|_{\varphi = \varphi_b} =0$, with the boundary conditions
\begin{equation}
\lim_{\tau\to\pm\infty}  \varphi(\tau, {\bf r})=\varphi_+
\label{bound1}
\end{equation}
and
\begin{equation}
\left.\frac{\partial\varphi (\tau,{\bf r})}{\partial\tau}\right|_{\tau=0}=0.
\label{bound2}
\end{equation}
The boundary conditions given by Eqs.~(\ref{bound1})--(\ref{bound2}) do not determine a unique solution for the bounce equation of motion. Consequently, in principle, all contributions in Eq.~(\ref{decayrate}) should be included. However, as shown by Coleman, Glaser, and Martin~\cite{martin}, the solution that minimizes $S_E$ (and hence provides the dominant contribution to the one-loop decay rate ) is a four-dimensional spherically symmetric configuration.
In $d$ dimensions the solution
has $O(d)$ symmetry.  In $d=4$, denoting $\rho^2=\tau^2+x^2+y^2+z^2$, then 
$\varphi_b$ follows the four-dimensional  radial equation of motion 
\begin{equation}
\frac{d^2\varphi_b}{d\rho^2}+ \frac{3}{\rho} \frac{d\varphi_b}{d\rho}=U'(\varphi_b) \;,
\label{b}
\end{equation} 
\noindent
with the boundary conditions 
\begin{equation}
\lim_{\rho\rightarrow
\infty}\varphi_b(\rho)=\varphi_+, ~~~ \frac{d\varphi_b}{d\rho}|_{\rho=0}=0
\label{sbound}
\end{equation}
which guarantee the existence of a unique solution
to  Eq. (\ref{b}) with finite Euclidean action.


From the above considerations, computing the one-loop contribution of $\Gamma/V$ requires, as a first step, solving a nonlinear equation for the bounce solution, followed by the evaluation of a nontrivial functional determinant that depends explicitly on this solution. 
Therefore, except for a few special cases \cite{shellard, guada1}, one typically has to rely either on approximate analytical approaches \cite{wipf, konoplich, strumia, flores, munster, ivanov, wen} or on numerical methods \cite{baacke, dunne, hur}. 
Nevertheless, the dominant contribution to the one-loop decay rate arises from the exponential factor. Consequently, to obtain an estimate of the tunnelling decay rate, it is common to approximate the exponential prefactor in Eq.~(\ref{decayrate}) by unity. 
Even with this simplification, the spherical bounce equation, Eq.~(\ref{b}), still lacks analytical solutions for most models \cite{fubini, duncan, dutta1, guada2, aravind, shen, lee, dutta2}. To overcome this difficulty, several efficient numerical techniques have been developed \cite{wainwright, masoumi, athron, sato, guada3}. 
In addition, a widely used approach in the literature is the so-called thin-wall approximation (TWA) method \cite{Coleman}, along with its subsequent refinements \cite{dunne}. The original TWA method, however, is restricted to situations in which the energy density difference between the true and false vacua is small compared with the barrier height of the potential. 

 The aim of this  letter is to introduce a variational framework that provides a systematic and robust alternative to the TWA, while remaining applicable well beyond its traditional domain of validity. We employs a simple but effective variational ansatz, allowing analytical control over the action 
and bubble profile even when the TWA assumptions break down.
Our study is motivated by the prevalence of metastable vacuum states in quantum field theories. For instance, in the 
Standard Model of particle 
physics the current Higgs vacuum may be metastable, prompting extensive research into its possible decay in the far future. More generally, many
 beyond-the-Standard-Model scenarios predict false vacua separated by significant potential barriers. Understanding the tunnelling between such 
vacua is crucial for assessing the fate of metastable phases in high-energy physics \cite{Andreassen2018}. Our variational approach  provides a  tool to analyze 
these decay processes beyond the limitations of the thin-wall approximation \cite{dunne,aravind}. Onother possible application is in 
first-order phase transitions in the early Universe \cite{Linde,Rubakov,Mark,Masina}
  where classical examples
 include transitions during inflation and cosmological phase transitions that can produce bubbles of true vacuum. The nucleation and growth of 
these bubbles are not only essential for phenomena like baryogenesis \cite{Yamada}, but can also generate  backgrounds of gravitational waves 
\cite{caprini1}. In this broader context, the accurate estimation of the tunneling rate beyond the thin-wall limit becomes essential, since cosmological 
potentials often exhibit large energy gaps between minima rather than nearly degenerate vacua. Therefore, analytical methods valid beyond the thin-wall regime are of phenomenological relevance.
\\

{\it The method:}
Since the spherical bounce corresponds to an extremum of the Euclidean action, rather than solving the nonlinear equation given by Eq.~(\ref{b}), we adopt an ansatz for the bounce in terms of free parameters, $\varphi_b = \varphi(\rho, a_1, a_2, \ldots)$. 
We then evaluate the Euclidean action for this ansatz, 
$S_E[\varphi(\rho, a_1, a_2, \ldots)] = S_E(a)$, 
and determine the parameters $a_i$ from the stationarity conditions
\begin{equation}
\frac{\partial}{\partial a_i} S_E(a) = 0.
\end{equation}
In this way, the original nonlinear differential equation is replaced by a system of algebraic equations. 

Two remarks are in order regarding this variational procedure. 
First, there is no {\it a priori} guarantee of its accuracy. To improve precision, the number of variational parameters $a_i$ must be increased, with the optimal choice determined by a compromise between analytical simplicity and the minimal requirements that the bounce ansatz must fulfill. 
Second, the spherical bounce is a saddle point, not a true minimum, of the functional $S_E(\varphi)$. The unstable direction in functional space, along which $S_E(\varphi)$ decreases at the saddle point, is associated with scale transformations $\rho \to \sigma \rho$. 
Accordingly, a variational parameter $\sigma$ must be included in the ansatz, leading to the form $\varphi_b = \varphi(\sigma \rho, a_1, a_2, \ldots)$.

To illustrate the method, we consider the $\varphi^4$ model with two non-degenerate minima, described by the potential density  
\begin{equation}
U(\varphi) = \lambda\left(\varphi^2 - \varphi_0^2 \right)^2 + \epsilon \varphi + U_0 \,,
\label{pot.1}
\end{equation}
where $\lambda > 0$ and $U_0$ is a constant chosen such that $U(\varphi_+) = 0$. 
Without loss of generality, we set $\varphi_0 = 1$, $\lambda = 1$, and $\epsilon > 0$. Other cases can be obtained through the rescaling of the field and coordinates, 
$\varphi \to -\varphi/\varphi_0$, $~x \to \tfrac{\sqrt{\lambda}}{\varphi_0} x$.  
In this parametrization, the potential exhibits two relative minima provided that $\epsilon$ lies in the interval
\[
0 < \epsilon < \left(\frac{2}{\sqrt{3}}\right)^3.
\]
The relative minimum $\varphi_+$ is determined from the condition $U'(\varphi) = 0$, which yields
\[
4\varphi_+^3 - 4\varphi_+ + \epsilon = 0.
\]

For the variational bounce we adopt the ansatz
\begin{equation}
\label{Bounce}
\varphi_b(\rho) = \varphi_+ + \alpha \exp\!\left[-(\sigma \rho)^{\beta}\right] \,,
\end{equation}
with three variational parameters, $\alpha$, $\beta$, and $\sigma$. 
 The variational parameters introduced in our ansatz have clear physical interpretations. In 
particular, $\alpha$ quantifies the difference in field value between the false vacuum $\varphi_+$ and the bounce at $\rho=0$, effectively 
measuring how deep into the true vacuum $\varphi_-$ the bounce solution reaches. The parameter $\beta$ controls the shape 
of the bounce interpoloting  the two vacua. Larger values of $\beta$ correspond to a steeper transition  between $\varphi_- $ 
and $\varphi_+$, while smaller $\beta$ yield a more gradual crossover. Finally, $\sigma$ sets the overall scale (size) of the bounce configuration. 
This scale parameter reflects, as mentioned above,  the fact that the bounce equation has an unstable direction associated with dilations.
By tuning $\alpha$, $\beta$, and $\sigma$, the ansatz can thus mimic 
key features of the exact bounce, with $\beta>1$ ensuring the correct boundary conditions (\ref{sbound}) and
 $\sigma$ accounting for the unstable scaling mode.

Substituting Eq.~(\ref{Bounce}) into Eq.~(\ref{action}), we obtain
\begin{equation}
S_E(\alpha,\beta,\sigma) = 2\pi^2 \Bigg\lbrace 
\dfrac{\alpha^2 \beta \sigma^{-2}}{8 (2)^{2/\beta}} \,
\Gamma\!\left(2 + \dfrac{2}{\beta}\right) 
+ \dfrac{\alpha^2 \sigma^{-4}}{\beta} \,
\Gamma\!\left(\dfrac{4}{\beta}\right)
\left[ \dfrac{\alpha^2}{4^{4/\beta}} + \dfrac{4\varphi_+\alpha}{3^{4/\beta}} 
+ \dfrac{2(3\varphi_+^{2}-1)}{2^{4/\beta}} \right] 
\Bigg\rbrace \,.
\label{actionvar}
\end{equation}

To simplify the computation, we first impose 
$\partial S_E / \partial \sigma = 0$ and $\partial S_E / \partial \alpha = 0$, 
from which we obtain
\begin{equation}
\sigma = \left\lbrace 
-\dfrac{16 (2)^{2/\beta} \Gamma\!\left(\tfrac{4}{\beta}\right)}{\beta^2 \Gamma\!\left(2 + \tfrac{2}{\beta}\right)}
\left[ \dfrac{\alpha^2}{4^{4/\beta}} + \dfrac{4\varphi_{+}\alpha}{3^{4/\beta}}
+ \dfrac{1}{2^{4/\beta}}\left( 3\varphi_{+}^{2} - 1 \right)  \right]
\right\rbrace^{1/2} ,
\end{equation}
and
\begin{equation}
\alpha = \dfrac{3^{4/\beta}}{2^{4/\beta}\varphi_+}\left(1 - 3\varphi_{+}^{2}\right)\,.
\end{equation}

By substituting these expressions into Eq.~(\ref{actionvar}), the Euclidean action reduces to
\begin{equation}
S_E(\beta) = 
\dfrac{\pi^2 (18)^{4/\beta} \beta^3 \, \Gamma^2\!\left(2 + \tfrac{2}{\beta}\right)
\left(3\varphi_{+}^{2} - 1\right)}{128 \,
\Gamma\!\left(\tfrac{4}{\beta}\right)
\Big[\big(2(8)^{4/\beta} - 3(3)^{8/\beta}\big)\varphi_{+}^{2} + 3^{8/\beta}\Big]} \,.
\end{equation}

Finally, the variational parameter $\beta$ is determined from the condition $\partial S_E / \partial \beta = 0$, which we solve for given values of $\epsilon$.

\begin{figure}[h]
\begin{center}
\includegraphics[width=0.5\linewidth]{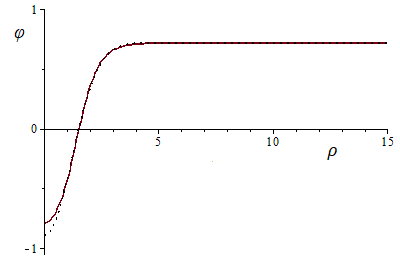}
\caption{Comparison between the variational bounce (dotted curve) and the numerical bounce (solid curve) for $\epsilon = 1.4$.}
\label{bounce1}
\end{center}
\end{figure}

As an illustration, for $\epsilon = 1.4$ we obtain $\beta = 2.1$ and $S_E = 56.7$, which is only 
$1.5\%$ larger than the corresponding numerical result, $S_E = 55.8$. 
In Fig.~\ref{bounce1} we show a comparison between the variational and numerical bounce solutions.  

\begin{figure}[h]
\centering
\includegraphics[width=0.7\linewidth]{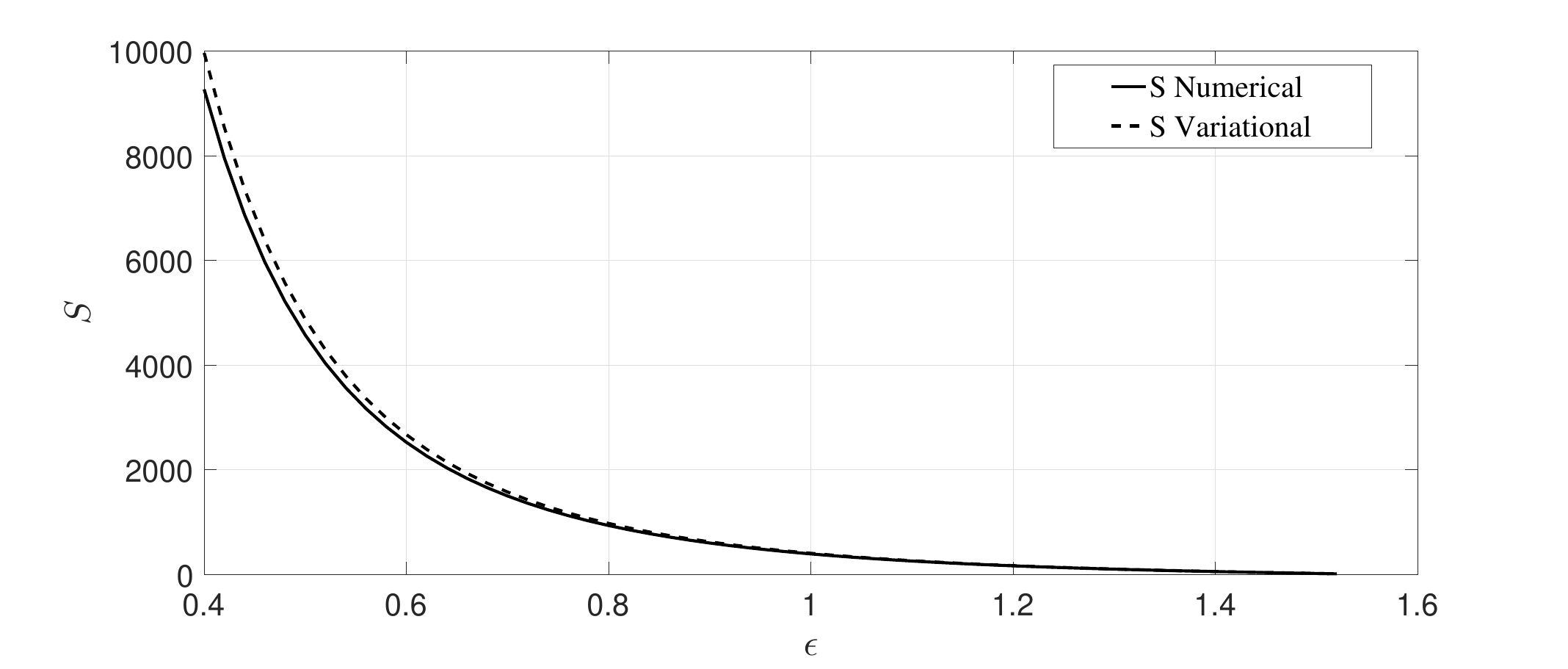}
\caption{Comparison between the Euclidean actions obtained from the variational approach and from numerical calculations as a function of $\epsilon$.}
\label{bounce2}
\end{figure}

For other values of $\epsilon$, the variational and numerical bounce solutions exhibit similarly close agreement. 
In Fig.~\ref{bounce2}, we present a direct comparison of the Euclidean actions obtained from the variational method and from numerical computations across a range of values of $\epsilon$.
\\

{\it Conclusions:}
In this work, we have demonstrated a strong agreement between the proposed variational method and numerical calculations over a broad range of $\epsilon$ values, as illustrated in Fig.~\ref{bounce2}. It is worth emphasizing that the variational bounce closely follows the numerical solution for large values of $\rho$, while noticeable deviations appear near the origin at $\rho = 0$ (see Fig.~\ref{bounce1}). This feature arises from the fact that the variational ansatz is constructed to satisfy the boundary condition at $\rho \to \infty$, where the approximation becomes exact. The accuracy of the method improves as $\epsilon$ increases, but gradually decreases for smaller values of $\epsilon$. Since the parameter $\epsilon$ quantifies the energy difference $U(\varphi_+) - U(\varphi_-)$, the TWA remains reliable in the regime of small $\epsilon$. Therefore, the present approach should be regarded as complementary to the TWA method: it remains valid beyond the thin-wall regime, while still exhibiting strong consistency with numerical solutions. 
 On qualitative grounds, we provided the physical interpretation of the variational parameters. In particular, the parameter $\alpha$ quantifies how deeply the bounce  penetrates into the true 
vacuum $\varphi_-$.

Looking ahead, we intend to extend this framework to the more
challenging problem of incorporating finite-temperature effects \cite{Linde}. 
 We also intend to explore applications in scenarios of current phenomenological \cite{Mark,Masina} and theoretical interest 
\cite{kumar1, dine1, dine2, ref1}.
 One immediate application is the inclusion of gravitational effects in false-vacuum decay. Incorporating gravity, as in the Coleman-De Luccia formalism
\cite{deLuccia,kanno1}, requires solving the bounce equation in curved spacetime, and a variational approach may simplify this task.  Furthermore, the variational strategy can be generalized to multi-field potentials  and metastable extended objects (such as topological defects), where traditional thin-wall approximations may fail  \cite{dunne,aravind}. These extensions are under current study and could broaden the scope of our method to a wide range of first-order transition phenomena.
\\
\\
{\bf Acknowledgements}\\
We gratefully acknowledge N.F. Svaiter for reading and providing valuable feedback that helped improve this paper.


\end{document}